\documentclass[aps,pre,amsmath,twocolumn,showpacs]{revtex4-1}
\usepackage{amsmath}
\usepackage{amsfonts}
\usepackage{graphicx}
\usepackage{hyperref}
\usepackage{mathtools}
\usepackage{verbatim}
\usepackage{epstopdf}
\usepackage{subfigure}
\usepackage{latexsym}
\usepackage{dcolumn}
\usepackage{epsf}
\usepackage{float}

\usepackage{color}
\begin{document}
\title{Transitions in overstable rotating magnetoconvection }
\author{Ankan Banerjee, Manojit Ghosh and Pinaki Pal*}
\affiliation{Department of Mathematics, National Institute of Technology, Durgapur 713209, India}
\date{\today}

\email{pinaki.math@gmail.com}

\begin{abstract}
The classical Rayleigh-B\'{e}nard convection (RBC) system is known to exhibit either  subcritical  or supercritical transition to convection in the presence or absence of rotation and/or magnetic field. However, the simultaneous exhibition of subcritical  and supercritical branches of convection in plane layer RBC depending on the initial conditions, has not been reported so far. Here, we report the phenomenon of simultaneous occurrence of subcritical  and supercritical branches of convection in overstable RBC of electrically conducting low Prandtl number fluids (liquid metals) in the presence of an external uniform horizontal magnetic field and rotation about the vertical axis. Extensive three dimensional (3D) direct numerical simulations (DNS) and low dimensional modeling of the system, performed in the ranges $750 \leq \mathrm{Ta} \leq 3000$ and  $0 < \mathrm{Q} \leq 1000$ of the Taylor number ($\mathrm{Ta}$, strength of the Coriolis force) and the Chandrasekhar number ($\mathrm{Q}$, strength of the Lorenz force) respectively, establish the phenomenon convincingly. Detailed bifurcation analysis of a simple three dimensional model derived from the DNS data reveals that a supercritical Hopf bifurcation and a subcritical pitchfork bifurcation of the conduction state are responsible for this.  The effect of Prandtl number on these transitions is also explored in detail.
\end{abstract}
%%%%%%%%%%%%%%%%%%%%%%%%%%%

%%%%%%%%%% Insert the texts which can accomdate on firstpage in the tag "fmtext" %%%%%
\maketitle
%\begin{fmtext}
\section{Introduction}
Overstable convection or overstability has drawn considerable attention of the researchers over the years due to its appearance in various astrophysical applications like the convective motion in sunspot and stellar interiors~\cite{danielson:1961, kato:1966, parker:1979, moore:1973, savage:1969, roberts:1976, lee_MNRA:1986, latter_MNRA:2015}. To understand the basic properties of overstable convection, researchers often consider simplified convection models like Rayleigh-B\'{e}nard convection (RBC)~\cite{eltayeb:1975, Roberts:2000, Jones:2000, podvigina_FD:2008, hirdesh:2013} in the presence of rotation and/or magnetic field. The overstable convection occurs in RBC as the `principle of exchange of stability' becomes invalid and the instability is manifested as a time dependent, oscillatory convective motion. The presence of external factors like magnetic field or/and rotation causes overstability since they provide an elastic-like behavior to the fluid so that it can sustain appropriate modes of wave propagation~\cite{vest:1969, moore:1973}. A comprehensive treatment on overstability using linear theory in the presence of rotation or/and magnetic field can be found in the classical monograph by Chandrasekhar~\cite{chandra:book}. 

Extensive theoretical as well as experimental investigations have been performed on overstability in the presence of rotation or external magnetic field over the last several decades\cite{goroff:1960, veronis:1959, knobloch:1990, clune:1993, boubnov:book, proctor:book}. As a result, a reasonable understanding on both linear and nonlinear aspects of the problem has been developed. However, the problem of overstable oscillatory convection in the presence of both rotation and magnetic field has received much less attention. In a rotating magnetoconvection (RMC) system, rotation introduces the Coriolis force, while magnetic field generates the Lorentz force. The presence of these two forces along with the buoyancy force make the problem more interesting even in its most simplified form like RBC. 

Overstability in such a simplified model of RMC, where an infinitely extended thin horizontal layer of electrically conducting fluid is heated uniformly from below in the presence of rotation and magnetic field was first theoretically investigated by Chandrasekhar~\cite{chandra:book}. The mathematical description of the problem consists of five non-dimensional parameters namely the Rayleigh number ($\mathrm{Ra}$, measures the vigor of buoyancy force), the Taylor number ($\mathrm{Ta}$, measures the rotation rate), the Chandrasekhar number ($\mathrm{Q}$, measures the strength of magnetic field), the Prandtl number ($\mathrm{Pr}$, the ratio of kinematic viscosity and thermal diffusivity of the fluid) and the magnetic Prandtl number ($\mathrm{Pm}$, the ratio of kinematic viscosity and magnetic diffusivity of the fluid). Chandrasekhar carried out extensive linear stability analysis of the conduction state to determine the onset of overstability by considering rotation about the vertical axis and vertical magnetic field. He obtained the critical Rayleigh number ($\mathrm{Ra_o}$) and wave number ($k_o$) for liquid metals ($\mathrm{Pr} = 0.025$, $\mathrm{Pm} \approx 0$) at the overstability onset corresponding to different $\mathrm{Ta}$ and $\mathrm{Q}$. Later, in two subsequent experimental works, Nakagawa~\cite{nakagawa:1957, nakagawa:1959} verified the theoretical findings of Chandrasekhar. Subsequently, Eltayeb~\cite{eltayeb:1975}  studied the overstable RMC for various orientations of the magnetic field and rotation with different types of boundary conditions by performing detailed asymptotic analysis. He determined some well defined scaling laws for the onset of overstability in the infinite $\mathrm{Ta}$ and $\mathrm{Q}$ limits.
Roberts and Jones~\cite{Roberts:2000, Jones:2000} theoretically investigated the RMC system in the presence of  a horizontal magnetic field and rotation about the vertical axis to determine the preferred mode of convection for very large Prandtl number fluids. Their linear analysis revealed the presence of different flow patterns including overstable cross rolls and overstable oblique rolls  at the onset. Later, Podvigina~\cite{podvigina_FD:2008} studied the RMC system theoretically with no-slip boundary conditions using linear theory to determine the parameter space where convective instability sets in as overstability. In a recent work, Eltayeb~\cite{Eltayeb:2013} performed linear stability analysis of a RMC system in the presence of  a horizontal magnetic field and rotation about the horizontal axis. The magnetic field and the axis of rotation were considered to be inclined at an angle $\phi$. The aim of the study was to understand the roles of viscosity, the electrical conductivity of the boundary and the interaction among all possible wave motions. 

It is evident from the literature that most of the investigations carried out in the field of overstability in the presence of rotation and magnetic field are based on linear theory. 
However, the nonlinear aspects of overstable convection of electrically conducting low Prandtl number fluids near the onset in the simultaneous presence of rotation and magnetic field have not been investigated yet. In this paper, we investigate different transitions  and associated bifurcation structures that occur close to the overstability onset by performing three dimensional (3D) direct numerical simulations (DNS) and low dimensional modeling of the RMC system in the presence of rotation about the vertical axis and a horizontal uniform magnetic field. We also explore the heat transfer properties of the system in detail.  The investigation has been performed here in the parameter ranges $0<\mathrm{Pr}\leq 0.5$, $750\leq\mathrm{Ta}\leq 3000$ and $0<\mathrm{Q}\leq 1000$ where overstability occurs at the onset. On the other hand, for a weaker rotation rate, the onset of convection is found to be always stationary. The transitions to convection including the bifurcation structure and pattern dynamics for such weaker rotation rate ($0< \mathrm{Ta}\leq 500$), where stationary convection occurs, have been discussed in detail in a recent study~\cite{ghosh_POF:2020}. The results of the investigation reported in the present study are mostly on low Prandtl number fluids (liquid metals) since they exhibit a very rich bifurcation structure near the onset of stationary as well as overstable convection in the absence or presence of magnetic field or rotation~\cite{busse:JFM_1972, jenkins_proctor:JFM_1984, meneguzzi:1987, clever:POF_1990, Baig:2008, nandu:2016, maity:EPL_2013, hirdesh:2013, nandu:2015, ghosh_POF:2020,mkv:book}. 

 \section{PHYSICAL SYSTEM AND LINEAR STABILITY ANALYSIS}\label{sec:rotating-hydromagnetic-systems}
We consider the classical Rayleigh-B\'{e}nard geometry in which an infinitely extended thin horizontal layer of electrically conducting fluid of thickness $d$, coefficient of thermal expansion $\alpha$, thermal diffusivity $\kappa$, kinematic viscosity $\nu$ and magnetic diffusivity $\lambda$ is confined between two horizontal plates. The plates are perfect conductors of heat and electricity. The bottom plate is heated uniformly and the top plate is kept cooler to maintain a steady adverse temperature gradient $\beta = \frac{\Delta T}{d} = \frac{T_l - T_u}{d}$ across the fluid layer, where $T_l$ and $T_u$ are temperatures of the top and bottom plates, respectively, with $T_l > T_u$. The system is rotated about the vertical axis with angular velocity $\Omega$ in the presence of a uniform external horizontal magnetic field $\mathbf{B_{0}}\equiv(0,B_0,0)$. The external magnetic field is attached to the system and co-rotates with it. For low Prandtl number fluids ($\mathrm{Pr} = 0.1$) the Froude number $\mathrm{Fr} = \frac{\Omega^2 L}{g}$ is always less than $2.6\times 10^{-4}$ for the rotation rates considered in this study ($750 \leq \mathrm{Ta} \leq 3000$). Thus, the effects of centrifugal force are neglected here. The stationary conduction state subjected to an external magnetic field in the rotating frame of reference is then considered as the basic state. The dimensionless system of equations which govern the convective flow of the system under the  Boussinesq approximation is given by

\begin{eqnarray}
\frac{\partial \bf{u}}{\partial t} + (\bf{u}.\nabla)\bf{u} &=& -\nabla{\pi} + \nabla^2{\bf{u}} + \mathrm{Ra} \theta {\bf{\hat{e_3}}}+ \sqrt{\mathrm{Ta}}(\bf{u}\times {\bf{\hat{e_3}}})\nonumber \\ && +\mathrm{Q}\left[\frac{\partial{\bf b}}{\partial y} + \mathrm{Pm}({\bf b}{\cdot}\nabla){\bf b}\right] , \label{eq:momentum} \\   
\mathrm{Pm}[\frac{\partial{\bf b}}{\partial t} + ({\bf u}{\cdot}\nabla){\bf b} &-& ({\bf b}{\cdot}\nabla){\bf u}] = {\nabla}^2 {\bf b} + \frac{\partial{\bf u}}{\partial y},\label{eq:magnetic}\\      
%\nabla^2\bf{b} &=& -\frac{\partial \bf{v}}{\partial y}, \label{magnetic} \\
\mathrm{Pr}\left[\frac{\partial \theta}{\partial t}+(\bf{u}.\nabla)\theta\right] &=& u_3+\nabla^2\theta \label{eq:heat},\\
\nabla . {\bf{u}}=0, && \nabla . {\bf{b}}=0. \label{eq:div_free}
\end{eqnarray}
In the above mathematical description $\textbf{u}(x,y,z,t)=(u_1,u_2,u_3)$ is the convective velocity field, $\theta(x,y,z,t)$ is the deviation in temperature field from steady conduction profile, $\pi(x,y,z,t)$ is the modified pressure field, $\textbf{b}(x,y,z,t)=(b_1,b_2,b_3)$ is the induced magnetic field and $\bf{\hat{e}_3}$ is the unit vector in vertical direction antiparallel to the gravitational acceleration $\bf{g}$. The non-dimensionalization procedure is accomplished by measuring all the length scales in the units of fluid thickness $d$, time scales in the units of viscous diffusion time scale $\frac{d^2}{\nu}$, the convective temperature field in the unit of $\frac{\beta d \nu}{\kappa}$, the convective velocity field in the unit of $\frac{\nu}{d}$ and the induced magnetic field in the unit of $\frac{B_0 \nu}{\lambda}$. The non-dimensionalization procedure gives rise to five dimensionless numbers namely the Rayleigh Number $\mathrm{Ra}=\frac{\alpha \beta g d^4}{\kappa \nu}$, the Taylor Number $\mathrm{Ta}=\frac{4 \Omega^2 d^4}{\nu^2}$, the Chandrasekhar Number $\mathrm{Q}=\frac{{B_0}^2 d^2}{\nu \lambda \rho_0}$, the Prandtl number $\mathrm{Pr}=\frac{\nu}{\kappa}$ and the magnetic Prandtl number $\mathrm{Pm}=\frac{\nu}{\lambda}$. In this work, our objective is to uncover the instabilities and associated bifurcation structures occurring near the overstability onset of low Prandtl number electrically conducting fluids (liquid metals, fluids present in the inner core of earth) for which the magnetic Prandtl number is very small ($\mathrm{Pm}\approx 10^{-6}$)~\cite{chandra:book,roberts_book,busse:1982,clever:1989}. So, for simplicity, we consider the asymptotic limit $\mathrm{Pm}\rightarrow 0$. In this limit, the equations (\ref{eq:momentum}) and (\ref{eq:magnetic}) become
\begin{eqnarray} 
\frac{\partial {\bf{u}}}{\partial t} + ({\bf{u}}.\nabla){\bf{u}} &=& -\nabla{\pi} + \nabla^2{{\bf{u}}} + \mathrm{Ra} \theta {\bf{\hat{e_3}}}\nonumber \\ && + \sqrt{\mathrm{Ta}}({\bf{u}}\times {\bf{\hat{e_3}}})+\mathrm{Q}\frac{\partial{\bf b}}{\partial y}, \label{eq:momentum1} \\   {\mathrm{and}~~~}
\nabla^2\bf{b} &=& -\frac{\partial \bf{u}}{\partial y}. \label{eq:magnetic1} 
\end{eqnarray}
The bounding surfaces located at $\mathrm{z}=0$ and $1$ are considered to be stress-free and perfect conductors of heat and electricity. This implies
\begin{eqnarray}
\frac{\partial u_1}{\partial z}=\frac{\partial u_2}{\partial z}=u_3=\theta=0 \label{bc01}
~~\mathrm{and}~~b_3=\frac{\partial b_1}{\partial z}=\frac{\partial b_2}{\partial z}=0. \label{bc02}
\end{eqnarray}
Periodic boundary conditions are assumed in the horizontal directions for all convective fields. Therefore, the equations (\ref{eq:heat})-(\ref{eq:magnetic1}) along with the boundary conditions (\ref{bc01}) represent the above described system mathematically. 

We now proceed to determine the conditions for overstability onset using linear theory~\cite{chandra:book}. We consider the linearised version of the above set of governing equations and follow a similar procedure to that described in~\cite{ghosh_POF:2020}. In the process, we consider the expression of $u_3$ in terms of normal mode as
\begin{equation}
u_3 = W(z)exp[i(k_{x}x + k_{y}y) + \sigma t], \nonumber 
\end{equation}
and obtain the equation
\begin{eqnarray} 
&&(D^2-k^2-\mathrm{Pr}\sigma)[\{(D^2-k^2)(D^2-k^2-\sigma)+\mathrm{Q}k_y^2\}^2 + \mathrm{Ta}D^2\nonumber \\ &&(D^2-k^2)]W =-\mathrm{Ra}k^2[(D^2-k^2)(D^2-k^2-\sigma) \nonumber \\ &&+ \mathrm{Q}k_y^2]W,
\label{eq:linear_stability_final_form}
\end{eqnarray}
where $k=\sqrt{{k_x}^2+{k_y}^2}$ is the horizontal wave number with $k_x$ and $k_y$ are the wave numbers along $x$ and $y$ directions respectively.

We choose a trial solution $W(z) = Asin(\pi z)$ which is compatible with the boundary conditions to get the following stability condition
\begin{eqnarray}
&&(\pi^2+k^2+\mathrm{Pr}\sigma)[\{(\pi^2+k^2)(\pi^2+k^2+\sigma) + \mathrm{Q}k_y^2\}^2+ \mathrm{Ta}\pi^2 \nonumber \\ &&(\pi^2+k^2)] = \mathrm{Ra}k^2[(\pi^2+k^2)(\pi^2+k^2+\sigma) + \mathrm{Q}k_y^2].
\label{eq:stability_condition}
\end{eqnarray}
To determine the conditions for overstability onset we put $\sigma = i\sigma_1$ in (\ref{eq:stability_condition}) and by comparing the real and imaginary parts we get the expressions for $\mathrm{Ra}$ and $\sigma_1$ as
\begin{widetext}
\begin{eqnarray}
\mathrm{Ra}(\mathrm{Ta},\mathrm{Q},\mathrm{Pr})=2\frac{\pi^2+k^2}{k^2}\{(\pi^2+k^2)^2+\mathrm{Q}{k_y}^2)\}\Big[\frac{(\pi^2+k^2)^2+\mathrm{Pr}^2\sigma_1^2}{(1-\mathrm{Pr})(\pi^2+k^2)^2-\mathrm{Q}{k_y}^2\mathrm{Pr}}\Big]	\label{eq:rao}
\end{eqnarray}
%\end{widetext}
and
%\begin{widetext}
\begin{eqnarray}
\mathrm{\sigma_1}(\mathrm{Ta},\mathrm{Q},\mathrm{Pr})=\bigg[\bigg(\frac{\pi^2\mathrm{Ta}}{\pi^2+k^2}\bigg)\bigg(\frac{(1-\mathrm{Pr})(\pi^2+k^2)^2-\mathrm{Q}{k_y}^2\mathrm{Pr}}{(1+\mathrm{Pr})(\pi^2+k^2)^2+\mathrm{Q}{k_y}^2\mathrm{Pr}} \bigg) -\bigg(\pi^2+k^2+\frac{\mathrm{Q}{k_y}^2}{\pi^2+k^2}\bigg)^2 \bigg]^{1/2}.\label{eq:omega1}
\end{eqnarray}
\end{widetext}

\begin{figure}[h]
\includegraphics[height=!, width = 0.5\textwidth]{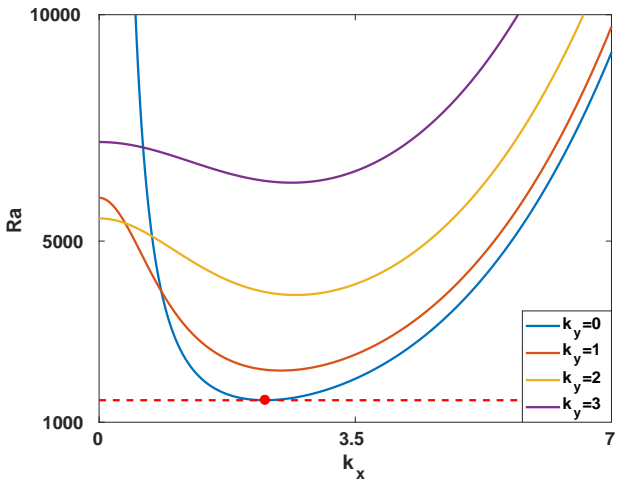}
\caption{Preferred mode of convection at the overstability onset computed using linear theory for $\mathrm{Pr} = 0.1$, $\mathrm{Ta} = 1100$ with $\mathrm{Q} = 100$ as a function of $k_x$ corresponding to different values of $k_y$.}
\label{fig:preferred_modes}
\end{figure}

\noindent From equations~(\ref{eq:rao}) and~(\ref{eq:omega1}), we notice that the Rayleigh number for overstability onset has complex dependency on  the parameters $\mathrm{Ta}$, $\mathrm{Q}$ and $\mathrm{Pr}$. Also, it is explicitly dependent on both $k_x$ and $k_y$. Therefore, we rely on numerical computation to determine the critical Rayleigh number for overstability onset ($\mathrm{Ra_o}$), corresponding critical wave number ($k_o$), associated angular frequency ($\sigma_1$) and the preferred mode of convection  corresponding to a fixed value of $\mathrm{Pr}$ using the expressions~(\ref{eq:rao}) and~(\ref{eq:omega1}) for given $\mathrm{Ta}$ and $\mathrm{Q}$. Fig.~\ref{fig:preferred_modes} shows the graphs of $\mathrm{Ra}$ (computed using  equations~(\ref{eq:rao}) and~(\ref{eq:omega1})) as a function of $k_x$ for different $k_y$ starting with $k_y = 0$ for $\mathrm{Pr} = 0.1$, $\mathrm{Ta} = 1100$ and $\mathrm{Q} = 100$. From the figure, we see that the minimum value of $\mathrm{Ra}$ occurs for $k_x = 2.27$ and $k_y = 0$. We obtain $\mathrm{Ra_o} = 1485.70$, $\sigma_1 = 19.12$ and $k_o = \sqrt{k_x^2 + k_y^2} = 2.27$ with $k_y = 0$ (solid blue curve in figure~\ref{fig:preferred_modes}). Therefore from linear theory we see that the preferred mode of convection is overstable two dimensional (2D) rolls for $\mathrm{Pr} = 0.1$, $\mathrm{Ta} = 1100$ with $\mathrm{Q} = 100$. We have checked that in the parameter ranges $0< \mathrm{Pr} \leq 0.5$, $750\leq \mathrm{Ta} \leq 3000$ and $0< \mathrm{Q} \leq 1000$, considered in this study, the preferred mode of convection is overstable 2D rolls for which $k_x \neq 0$ and $k_y = 0$. Thus, the instability is independent of $\mathrm{Q}$.
Also, the minimum value of $\mathrm{Ta}$ ($\mathrm{Ta_c}$) required for overstable oscillatory convection corresponding to a fixed value of $\mathrm{Pr}$ is independent of $\mathrm{Q}$. Further, the value of $\mathrm{Ta_c}$ grows rapidly as $\mathrm{Pr}$ increases and the scenario of overstability vanishes for $\mathrm{Pr} \geq 0.6766$~\cite{chandra:book}. Now, we proceed for direct numerical simulations (DNS) of the system to verify the results obtained from linear theory, details of which are discussed in the following section.

\section{Direct numerical simulations (DNS)}\label{sec:dns}
An object oriented pseudo-spectral code TARANG~\cite{mkv:code} is used to carry out DNS of the governing equations~(\ref{eq:heat})-(\ref{eq:magnetic1}) together with the boundary conditions (\ref{bc01}). Equation~(\ref{eq:magnetic1}) shows that the induced magnetic field is slaved to the velocity field. In the simulation code, the independent variables present in the governing equations i.e. vertical velocity, vertical vorticity and the deviation in temperature field are expanded using a set of orthogonal basis functions compatible with the boundary conditions as
\begin{figure}[h]
\includegraphics[height=!, width = 0.5\textwidth]{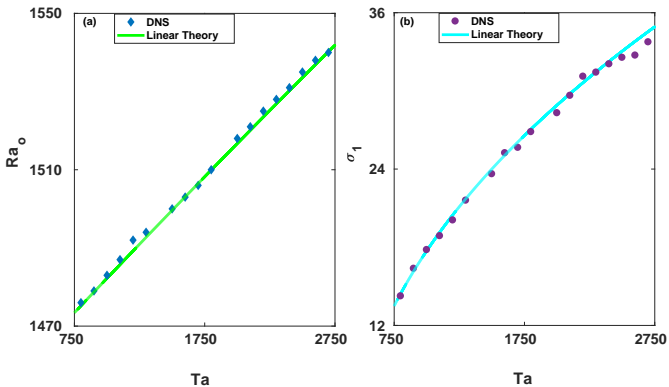}
\caption{Critical Rayleigh number ($\mathrm{Ra}_o$) and angular frequency ($\sigma_1$) at the onset of overstable convection are shown as a function of $\mathrm{Ta}$ for $\mathrm{Pr} = 0.1$. The green (for $\mathrm{Ra}_o$ in (a)) and cyan (for $\sigma_1$ in (b)) curves are obtained from the linear theory, while solid blue diamonds (represents $\mathrm{Ra}_o$ in (a)) and solid dots (represents $\sigma_1$ in (b)) are DNS data.
}
\label{fig:dns_validation}
\end{figure}

\begin{eqnarray}
u_3(x,y,z,t)&=&\sum\limits_{l,m,n} W_{lmn}(t)e^{i(lk_x x+mk_y y)}\sin( n \pi z), \nonumber\\
\omega_3(x,y,z,t)&=&\sum\limits_{l,m,n} Z_{lmn}(t)e^{i(lk_x x+mk_y y)}\cos( n \pi z), \nonumber\\
\theta(x,y,z,t)&=&\sum\limits_{l,m,n} \Theta_{lmn}(t)e^{i(lk_x x+mk_y y)}\sin( n \pi z). \label{eq:fourier_expansions}
\end{eqnarray}
The coefficients $W_{lmn}$, $Z_{lmn}$ and $\Theta_{lmn}$ are the Fourier coefficients and $l$, $m$ and $n$ can take any non-negative integer values including zero. $k_x$ and $k_y$ are the wave numbers along the $x$-direction and $y$-direction respectively. We set $k_x = k_y = k_o$ for  the present simulations. The horizontal components of the velocity and induced magnetic field are then derived by using the equation of continuity and equation~(\ref{eq:magnetic1}). Simulations are performed in a square box of size $(2\pi/k_o)\times(2\pi/k_o)\times1$ with spatial grid resolution $32^3$. Fourth order Runge-Kutta method is used for time advancement with time step $\delta t = 0.001$. Random initial conditions are used for the simulations.  We introduce a new parameter $r = \mathrm{Ra}/\mathrm{Ra_o}(\mathrm{Ta, Pr})$, called the reduced Rayleigh number in  the subsequent discussion. 

 Numerical investigation is carried out near the onset of convection over the parameter ranges $750\leq \mathrm{Ta} \leq 3000$, $0< \mathrm{Q}\leq 1000$ and $0<\mathrm{Pr}\leq 0.5$.  We first determine $\mathrm{Ra_o}$ from DNS for different values of $\mathrm{Ta}$ in the considered parameter range corresponding to $\mathrm{Pr} = 0.1$ using the values of $k_o$ obtained from the linear theory. The variation of $\mathrm{Ra_o}$ and associated $\sigma_1$ for different values of $\mathrm{Ta}$ obtained from linear theory and DNS are shown in figure~\ref{fig:dns_validation}. From the figure, it is clear that at the onset of overstability, the linear theory and DNS have a good agreement. Now, using the code we have performed extensive simulations in our considered parameter space to unfold different flow patterns which are discussed in Section~\ref{sec:results}.

\section{\label{sec:results}	RESULTS AND DISCUSSION}

\subsection{Effect of large magnetic field ($\mathrm{Q} \geq 100$)}\label{large_Q}
\subsubsection{DNS results}\label{DNS_result}
\begin{figure}[h]
\begin{center}
\includegraphics[height=2.7in, width = 0.5\textwidth]{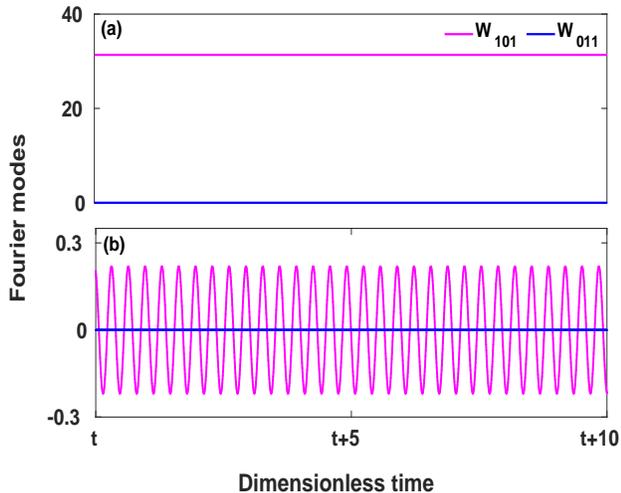}
\end{center}
\caption{Temporal variation of the Fourier modes  $W_{101}$ and $W_{011}$ near the onset of primary instability ($r=1.001$) for $\mathrm{Ta}=1100$, $\mathrm{Q}=100$ and $\mathrm{Pr}=0.1$ as obtained from DNS corresponding to (a) 2D rolls  and (b)  oscillatory rolls solutions.}
\label{fig:onset_ts}
\end{figure}

We perform extensive DNS in the considered ranges of $\mathrm{Ta}$ and $\mathrm{Q}$ for $\mathrm{Pr} = 0.1$. We first explore the effect of large magnetic field ($\mathrm{Q} \geq 100$) near the onset of overstable rotating convection. We observe multiple solutions at the onset of convection ($r = 1.001$) corresponding to different sets of initial conditions. Figure~\ref{fig:onset_ts} shows the existence of two different classes of solutions at the onset of convection for $\mathrm{Ta} = 1100$ and $\mathrm{Q} = 100$. A high amplitude 2D rolls solution for which $W_{101} \neq 0$ and $W_{011} = 0$ (see fig.~\ref{fig:onset_ts}(a)) appears at the onset corresponding to a different set of initial conditions along with the usual periodic oscillatory rolls solution for which $W_{101} \neq 0$, $W_{011} = 0$ and $W_{101}$ oscillates over time (see fig.~\ref{fig:onset_ts}(b)). Changes in $\mathrm{Ta}$ do not alter the scenario at the onset in our considered parameter range. However, changes in $\mathrm{Pr}$ have nontrivial effects which we will discuss later.   

\begin{figure}[]
\begin{center}
\includegraphics[height=2.7in, width = 0.5\textwidth]{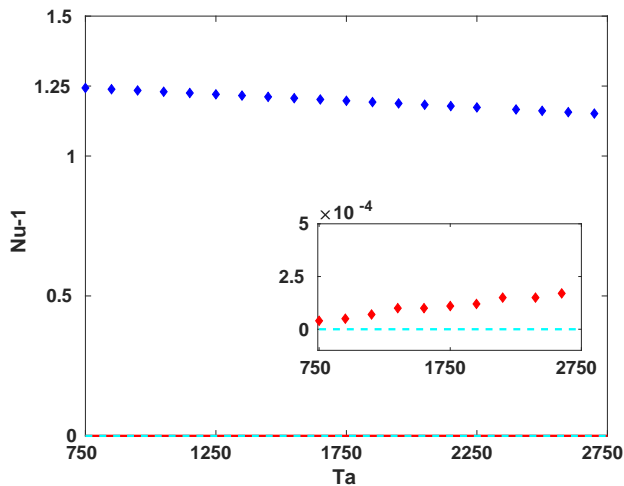}
\end{center}
\caption{Variation of convective heat flux $(\mathrm{Nu}-1)$ near the onset of primary instability ($r=1.001$) for $\mathrm{Q}=100$ and $\mathrm{Pr}=0.1$ corresponding to the 2D rolls solution (blue) and periodic solution (red) as a function of $\mathrm{Ta}$.  The variation of $(\mathrm{Nu}-1)$ corresponding to the periodic solution (red) is clearly visible at the inset. The cyan curve represents the straight line $\mathrm{Nu} = 1$ corresponding to the steady conduction state.}
\label{fig:onset_nu}
\end{figure}

The existence of multiple solutions at the onset of convection motivates us to study the heat transport properties associated with different solutions. Therefore, we compute the Nusselt number ($\mathrm{Nu}$, ratio of total heat flux to conductive heat flux) at the onset of convection corresponding to different solutions. The variation of convective heat flux ($\mathrm{Nu} - 1$) at the onset  corresponding to two different types of solutions for $\mathrm{Q} = 100$ is shown in figure~\ref{fig:onset_nu}. Interestingly, from the figure, we observe a sharp jump in $\mathrm{Nu}$ for the 2D rolls solution indicating a sudden enhancement in heat transfer. However, the $\mathrm{Nu}$ corresponding to the periodic oscillatory rolls solution shows a smooth transition. We also notice that variation in $\mathrm{Q}$ for fixed $\mathrm{Ta}$ has only a trivial effect on heat transport at the onset though heat transport at the onset depends on $\mathrm{Ta}$ for fixed $\mathrm{Q}$.
Figure~\ref{fig:onset_nu} also shows that $\mathrm{Nu}$ at the onset corresponding to 2D rolls solution decreases and that corresponding to periodic oscillatory rolls increases with $\mathrm{Ta}$. This surprising behavior of $\mathrm{Nu}$ hints at the possibility of different transitions to convection. 

To explore the possible occurrence of different transitions to convection at the onset we  
\begin{figure}[h]
\includegraphics[height=!, width = 0.5\textwidth]{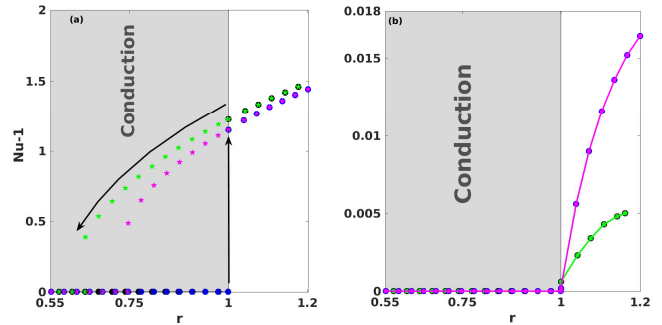}
\caption{Convective heat flux computed from DNS as a function of $r$ for $\mathrm{Q}=100$, $\mathrm{Pr} = 0.1$  and two different values of $\mathrm{Ta}$. Filled circles (black for $\mathrm{Ta} = 1100$ and blue for $\mathrm{Ta} = 2700$) and stars 
(green for $\mathrm{Ta} = 1100$ and pink for $\mathrm{Ta} = 2700$), respectively, represent the convective heat flux during forward and backward continuation for 2D rolls solutions in (a)  and  oscillatory rolls solutions in (b).}
\label{fig:Nu_all}
\end{figure}
perform forward and backward numerical continuation of the solutions observed in DNS for given $\mathrm{Ta}$ and $\mathrm{Q}$. We start with $r = 0.55$ and increase $r$ in small steps ($\Delta r = 0.03$) to $r = 1.2$ for the forward continuation. We use random initial conditions each time. On the contrary, for the backward continuation, we first simulate the system for $r = 1.2$ using random initial conditions. Then using the final results of last simulation as the current initial conditions we proceed for the present simulation by reducing $r$ in small steps ($\Delta r = 0.03$) up to $r = 0.55$. 

Observing the results of forward and backward continuation, we discover the simultaneous occurrence of  subcritical and supercritical branches of convection at the onset. Figures~\ref{fig:Nu_all}(a) and \ref{fig:Nu_all}(b) show the variation of $\mathrm{Nu} - 1$, close to the onset of convection as a function of $r$ obtained from DNS for forward and backward continuation. The variation of $\mathrm{Nu} - 1$ for 2D rolls solution shows a finite jump at $r = 1$ and follows different paths during forward and backward continuation (see fig~\ref{fig:Nu_all}(a)). Subsequently, a hysteresis loop appears and convection continues to exist in the conduction region ($r < 1$). A typical scenario of  subcritical transition prevails at the onset which is common to liquid-gas transitions, solid-liquid transitions, superconductors, percolation theory and many other fields~\cite{halperin:1974, gunton:1983, binder:1987, kuwahara:1995, parshani:2010, goldenfeld:2018, schrieffer:2018}. Note that, this scenario of  subcritical transition is independent of $\mathrm{Q}$ and solely depends on $\mathrm{Ta}$ for fixed $\mathrm{Pr}$. Also, from figure~\ref{fig:Nu_all}(a), we notice that the width of the hysteresis loop decreases as the value of $\mathrm{Ta}$ is increased. A scenario of  supercritical transition appears close to the onset of convection for periodic oscillatory rolls solution (see fig~\ref{fig:Nu_all}(b)). We neither observe a jump in $\mathrm{Nu} - 1$ nor a hysteresis in this case.

\begin{figure*}
\includegraphics[height=4.5cm, width=\textwidth]{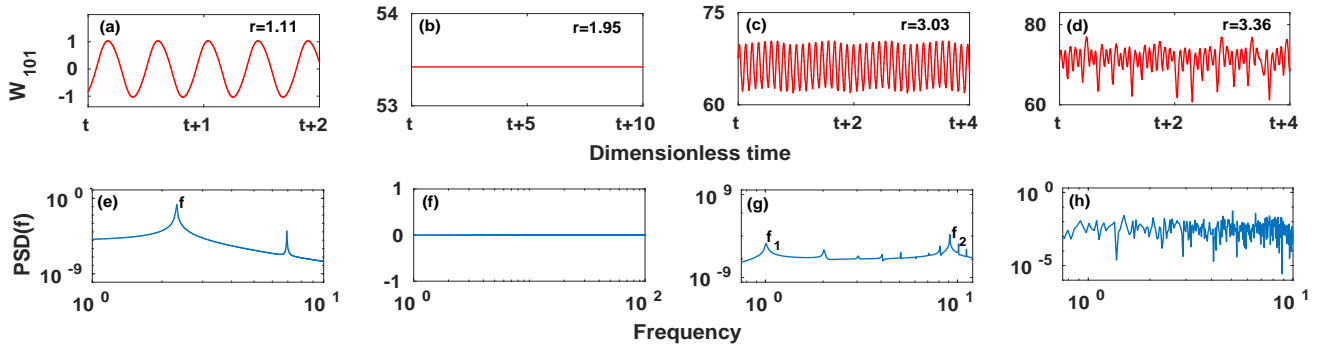}
\caption{First panel (a)-(d) displays time evolution of $W_{101}$ corresponding to the subsequent transitions of convective flow originated by overstable convection for $\mathrm{Ta}=1100$, $\mathrm{Q}=100$ and $\mathrm{Pr} = 0.1$. Second panel (e)-(h) displays corresponding power spectral density.}
\label{fig:transitions}
\end{figure*}

We also perform DNS by varying $r$ for fixed $\mathrm{Ta}$, $\mathrm{Q}$ and $\mathrm{Pr}$ to uncover the subsequent transitions after convection sets in. Figure~\ref{fig:transitions} shows the time evolution of the largest Fourier mode $W_{101}$ along with the power spectral density (PSD) corresponding to the subsequent transitions for $\mathrm{Ta} = 1100$, $\mathrm{Q} = 100$ with $\mathrm{Pr} = 0.1$ as $r$ is increased. From the figure, we see that very close to the onset small amplitude periodic oscillatory rolls persist (see fig.~\ref{fig:transitions}(a)). It vanishes as $r$ is raised and high amplitude 2D rolls appear (see fig.~\ref{fig:transitions}(b)). Further increment in $r$ brings the high amplitude quasiperiodic oscillatory rolls (see fig.~\ref{fig:transitions}(c)) followed by  the chaotic oscillatory rolls (see fig.~\ref{fig:transitions}(d)). Changes in $\mathrm{Ta}$ for fixed $\mathrm{Q}$ and $\mathrm{Pr}$ do not alter the scenario qualitatively. However, for fixed $\mathrm{Ta}$ and $\mathrm{Pr}$, modification in $\mathrm{Q}$ influences the flow patterns deeply. Table~\ref{table:effect_large_Q} shows the effect of $\mathrm{Q}$ on the flow structures for $\mathrm{Ta} = 1100$ and $\mathrm{Pr} = 0.1$. From the table we notice the high amplitude periodic solutions which appear following 2D rolls, now exist for much higher values of $r$ with the increment in $\mathrm{Q}$. As a result, the stability region of 2D rolls is increased and flow becomes two dimensional there. 
\begin{widetext}
\begin{center}
\begin{table}[h]
\caption{Effect of $\mathrm{Q}$ on convective flow patterns for $\mathrm{Ta} = 1100$ and $\mathrm{Pr} = 0.1$. These are the observations from DNS with the increment in $r$ starting with $r = 1$. Ranges are given for the instabilities having prime interest. Onset of the higher order instabilities are also included here.}\label{table:effect_large_Q}
\begin{tabular}{c|c|c| c | c}
\hline 
\hline
$\mathrm{Q}$ & $\mathrm{Oscillatory~Rolls}$ & $\mathrm{2D~Rolls}$ & $\mathrm{Quasi~Periodic~Rolls}$ & $\mathrm{Chaotic ~Rolls}$\\ 
\hline 
100 & 1 - 1.159 & 1 - 2.311 & 2.312 & 3.015 \\ 
500 & 1 - 1.206 & 1 - 7.891 & 7.892 & -- \\ 
1000 & 1 - 1.174 & 1 - 18.079 & 18.08 & --  \\
\hline
\hline
\end{tabular} 
\end{table}
\end{center}
\end{widetext}

From the above study, we see that DNS exhibits numerous stationary and time dependent solutions near the onset including simultaneous occurrence of subcritical and supercritical branches of convection. Also, the appearance of subcritical transition causes a substantial enhancement in heat transport near the onset. However, describing the underlying bifurcations and origins of these transitions using DNS is quite laborious. Therefore, we follow the low dimensional modeling technique and try to uncover the origin of different solutions and transitions observed in DNS by performing a bifurcation analysis. Next, we discuss the construction of the low dimensional model.

\subsubsection{A low dimensional model}\label{ldm}
We now derive a low dimensional model containing the minimum number of equations which can capture the simultaneous occurrence of  subcritical and supercritical branches of convection at the onset following the procedure described in~\cite{nandu:2016}. The key concept underlying the procedure is to identify the large scale modes present in DNS data by calculating the contribution of an individual mode to the total energy. Following the method, we identify only one vertical velocity mode: $W_{101}$, two vertical vorticity modes: $Z_{101}$, $Z_{200}$ and two modes in the temperature fluctuation: $\Theta_{101}$, $\Theta_{002}$. Therefore, the truncated expressions for $u_3$, $\omega_3$ and $\theta$ become
\begin{eqnarray}
u_3 &=& W_{101}(t)\cos k_o x \sin \pi z,\nonumber\\
\omega_3 &=& Z_{101}(t)\cos k_o x \cos \pi z + Z_{200}(t)\cos 2 k_o x,\nonumber\\
\theta &=& \Theta_{101}(t)\cos k_o x \sin \pi z + \Theta_{002}(t) \sin 2 \pi z. \label{model_expression}
\end{eqnarray}

Selection of the above five large scale modes from the DNS data can also be well understood from a theoretical perspective. Linear theory suggests that it is the mode $W_{101}$ in vertical velocity whose temporal growth rate first becomes zero at the onset of convection ($r = 1.001$). All the other modes present in vertical velocity have negative temporal growth rate there. Therefore, we choose only $W_{101}$ in the truncated expression of $u_3$. From equation~(\ref{eq:momentum1}) we see that the vertical vorticity couples linearly with the vertical velocity in presence of rotation ($\mathrm{Ta} \neq 0$). Hence we consider the mode $Z_{101}$ in the truncated expression of $\omega_3$. Also, from equation~(\ref{eq:heat}) we observe that $\theta$ and $u_3$ are linearly coupled which demands inclusion of the mode $\Theta_{101}$ in the truncated expression for $\theta$. 

Note that the $W_{101}$ mode physically represents the 2D rolls pattern along $y$-axis and the amplitude of $W_{101}$ starts to grow in time after convection sets in. However, as soon as the amplitude of $W_{101}$ becomes significant, the nonlinearity present in the system starts to influence the flow and generates higher Fourier modes. Due to this, many things appear in the system such as saturation in the growth rate of the primary mode, occurrence of stationary and time dependent patterns, chaos etc. Since from DNS, we have already seen the existence of both stationary and time dependent solutions at the onset, we now look for the minimal nonlinear interaction in order to include the effect of nonlinearity present in the system. 

Therefore, we proceed to the simplest nonlinear correction in the expressions of $u_3$, $\omega_3$ and $\theta$ effected by the nonlinear terms $(\bf{u}.\nabla)\bf{u}$ and $(\bf{u}.\nabla)\theta$ present in the momentum and energy equations respectively. The nonlinear correction generates the modes $Z_{200}$ in $\omega_3$ and $\Theta_{002}$ in $\theta$ through the triad interaction. Finally, we get the above truncated expressions for $u_3$, $\omega_3$ and $\theta$. 

Horizontal components of the velocity then can be easily found by using the expressions of $u_3$, $\omega_3$ and the continuity equation. We project the hydrodynamic system~(\ref{eq:heat})-(\ref{eq:magnetic1}) on these modes to get $5$ coupled nonlinear ordinary differential equations. We observe that the linear decay rate of $Z_{200}$ and $\Theta_{002}$ is much larger than that of $Z_{101}$ and $\Theta_{101}$. Therefore, the slow modes $Z_{101}$ and $\Theta_{101}$ together with $W_{101}$ drive the evolution of fast modes $Z_{200}$ and $\Theta_{002}$. In other words, the stable modes $Z_{200}$ and $\Theta_{002}$ become slaved to the unstable modes $W_{101}$, $Z_{101}$ and $\Theta_{101}$. We then eliminate the equations for the modes $Z_{200}$ and $\Theta_{002}$ adiabatically~\cite{mannevile:book_1990}. Finally, we arrive at a small system consisting of only three nonlinear ordinary differential equations, which is our desired low dimensional model given by
\begin{eqnarray}
\dot{X}&=&aX+bY+cZ,\nonumber \\
\dot{Y}&=&abX+aY-dX^2Y,\nonumber \\
\dot{Z}&=&\frac{a}{Pr} Z+\frac{1}{Pr} X-\frac{Pr}{8} X^2Z.
\label{model}
\end{eqnarray}
In the above system $X=W_{101}$, $Y=Z_{101}$, $Z=\Theta_{101}$ and the coefficients are $a=-(\pi^2+k_o^2)$, $b=-\frac{\pi \sqrt{\mathrm{Ta}}}{\pi^2+k_o^2}$, $c=\frac{\mathrm{Ra}k_o^2}{\pi^2+k_o^2}$, and $d=\frac{\pi^2}{8k_o^2}$ where $k_o$ is the critical wave number for the onset of overstability.

\subsubsection{Bifurcation Analysis}\label{bifurcation_analysis}
\begin{figure}[h]
\includegraphics[height=!, width = 0.5\textwidth]{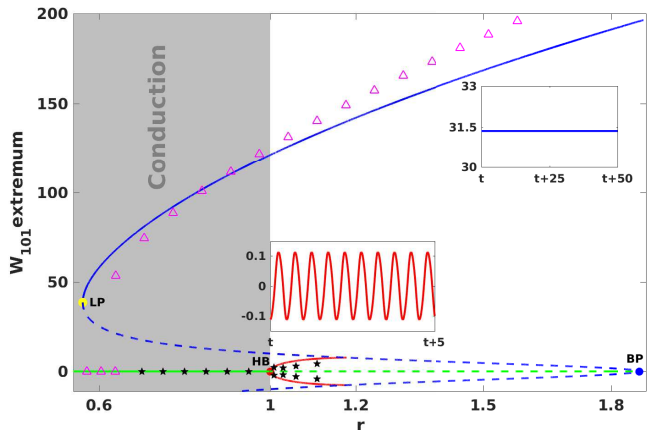}
\caption{Bifurcation diagram as obtained from the model for $\mathrm{Ta}=1100,~\mathrm{Q}=100$ and $\mathrm{Pr=0.1}$ . The stable and unstable solutions are represented by the solid and dashed lines respectively. The gray shaded region represents the conduction zone and the   green curves show the trivial conduction state. The red filled circle at $r=1$ shows the supercritical Hopf bifurcation point. Extremum values of the limit cycles are represented by the red curves. The dashed blue curves originated at the branch point BP at $r=1.88$ (filled blue circle) through a subcritical pitchfork bifurcation of the unstable conduction branch represent unstable 2D rolls solutions which becomes stable via a saddle node bifurcation near $\mathrm{r}=0.6$ (filled  yellow circle). The 2D rolls branch then turns towards higher $r$ and continue to exist as a stable solution (solid blue curve).  Both time dependent (supercritical origin) and finite amplitude steady (subcritical origin) solutions persist at the convection onset. Empty pink triangles and   black stars represent the data obtained from DNS for 2D rolls and periodic  oscillatory rolls solutions respectively. Insets show the time evolution of $W_{101}$ for a typical steady 2D rolls solution (blue) and a periodic  oscillatory rolls solution (red). }
\label{fig:bfd1100}
\end{figure}

We perform detailed bifurcation analysis of the model~(\ref{model}) using MATLAB based continuation software named MATCONT~\cite{dhooge:matcont_2003}. From the discussion in section~(\ref{DNS_result}) we note that the simultaneous appearance of subcritical and supercritical branches does not depend on $\mathrm{Ta}$ and $\mathrm{Q} ~(\geq 100)$ for fixed $\mathrm{Pr}$. Therefore, we prepare only one bifurcation diagram to explore the origin of simultaneous transitions occurring at the onset. Figure~\ref{fig:bfd1100} shows the bifurcation diagram constructed using the model for $\mathrm{Ta} = 1100$, $\mathrm{Q} = 100$ with $\mathrm{Pr} = 0.1$. Extremum values of $W_{101}$ corresponding to different solutions are displayed in the figure as a function of $r$ in the range $0.54 \leq r \leq 1.88$. Solid and dashed green curves represent the stable and unstable conduction solutions respectively. The stable conduction solution loses its stability at $r = 1$ via a supercritical Hopf bifurcation.
\begin{figure*}[]
\includegraphics[height=3in, width = \textwidth]{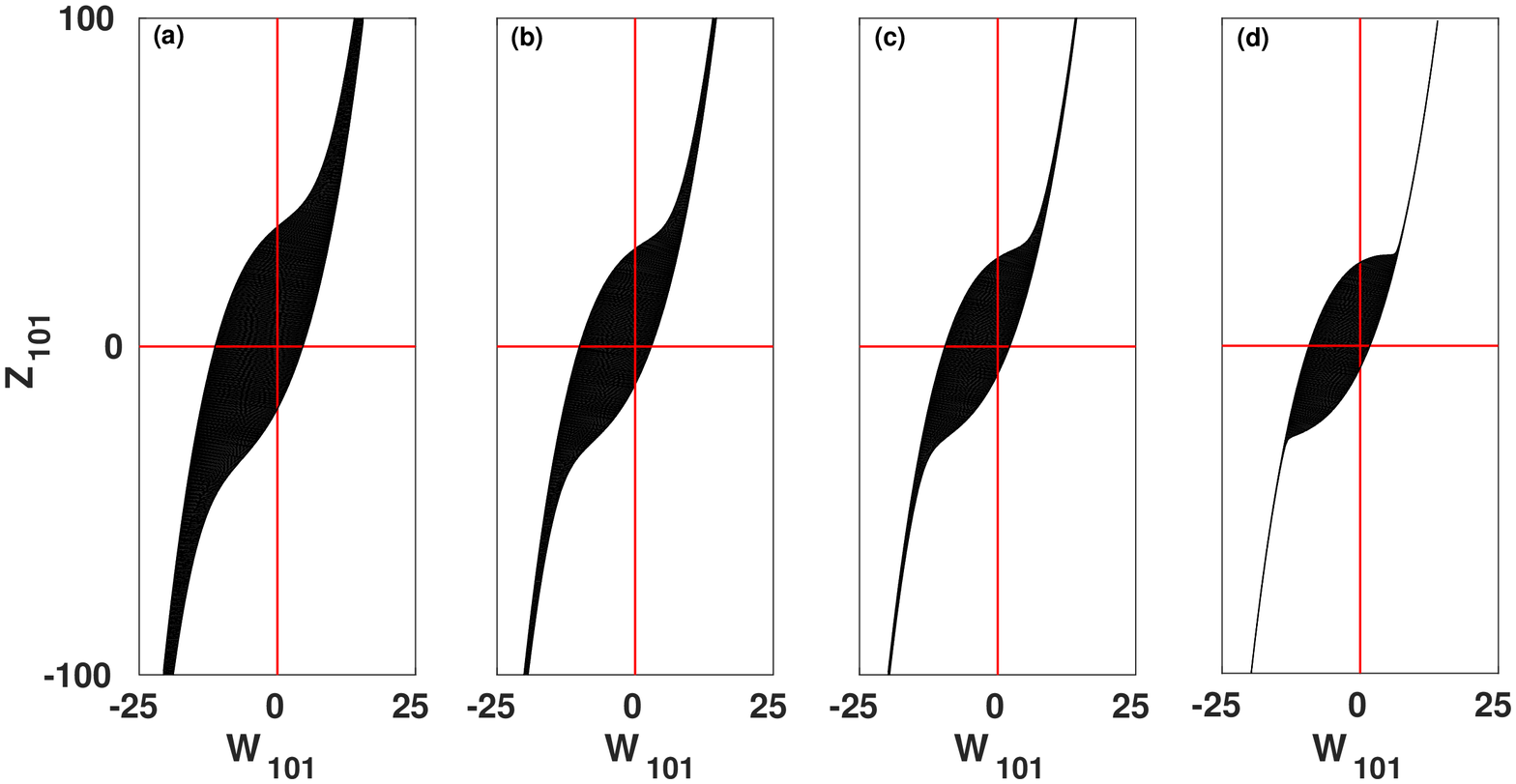}
\caption{Basins of attraction for steady 2D rolls and  oscillatory rolls solutions are shown on $\mathrm{W}_{101}-\mathrm{Z}_{101}$ plane for $\mathrm{Ta}=1100,~\mathrm{Q=100}$ and $\mathrm{Pr}=0.1$ as the reduced Rayleigh number is varied. Initial conditions from black (white) region lead to   oscillatory rolls solutions of supercritical origin (finite amplitude 2D rolls solution of subcritical origin). (a), (b), (c) and (d) are corresponding to the reduced Rayleigh numbers $\mathrm{r} = 1.0086$, $1.0758$, $1.1095$ and $1.1283$ respectively.}
\label{fig:basin_attraction}
\end{figure*}
The Hopf bifurcation point is shown with a filled red circle in the figure. Stable limit cycles appear due to this Hopf bifurcation. Note that, the eigenvectors at the Hopf bifurcation are for certain values of $W_{101}$, $Z_{101}$, $\Theta_{101}$ with $Z_{200} = \Theta_{002} = 0$. Extremum values of these limit cycles are displayed with solid red curves in the figure. Time evolution of the $W_{101}$ mode corresponding to these limit cycles is shown in the inset (solid red curve varying periodically with dimensionless time). The pattern dynamics of these limit cycles is similar to that reported in Ref.~\cite{busse:JFM_1972, pal:2012}.

The unstable conduction solution continues to exist for higher values of $r$ and goes through a subcritical pitchfork bifurcation at $r = 1.866$ (filled   blue circle). An unstable 2D rolls branch for which $W_{101} \neq 0$ and $W_{011} = 0$ is originated there (dashed blue curves). This unstable 2D rolls branch starts to move backward and continues to exist for lower values of $r$, even for $r < 1$. The unstable 2D rolls branch becomes stable via a saddle node bifurcation at $r = 0.56$ (filled yellow circle) inside the conduction region. The stable 2D rolls branch (solid blue curve) then changes its direction and continues to exist for higher values of $r$. This branch eventually comes out of the conduction region at $r = 1$. As a result, a high amplitude 2D rolls solution prevails at the onset which causes a sudden enhancement in heat transport there during the forward transition. The variation of $W_{101}$ with time for the 2D rolls solution is also shown in the inset.

\begin{figure*}[]
\includegraphics[height=!, width = \textwidth]{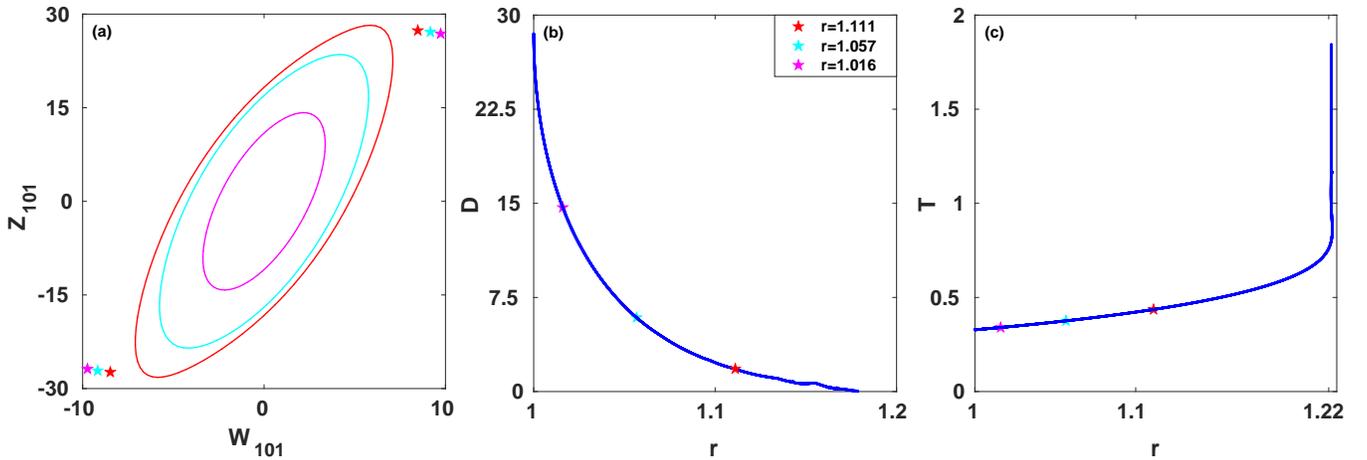}
\caption{Homoclinic bifurcation of the limit cycle generated via supercritical Hopf bifurcation (overstability) for $\mathrm{Pr} = 0.1$, $\mathrm{Ta} = 1100$, and $\mathrm{Q} = 100$. (a) Projection of the limit cycles on the $W_{101} - Z_{101}$ plane together with the 2D rolls saddle are shown corresponding to three different $r$. The increase in the size of the limit cycles is apparent as $r$ approaches the homoclinic bifurcation point $r \sim 1.22$.  (b) Distance $\mathrm{D}$ of the limit cycles from the 2D rolls saddle as a function of  $r$. The points for which limit cycles are shown in (a) are indicated with same color code. (c) Variation of time period of the limit cycle with $r$.  }
\label{fig:homoclinic_bif}
\end{figure*}

From the bifurcation diagram, we observe that  the stable 2D rolls solution coexists along with the stable conduction state inside the conduction region. As a result,  the 2D rolls solution continues to exist during backward continuation and convection persists in the conduction region. A  typical scenario of  subcritical transition accompanied with a hysteresis loop appears at the onset. The Hopf bifurcation point and the saddle node bifurcation point are the forward and backward transition points here respectively. The distance between these two points represents the hysteresis width. Results obtained from DNS also show good qualitative agreement with the model results. Empty pink triangles represent the DNS data corresponding to the 2D rolls solution for the specified parameter values during the backward transition. The black stars in the figure~\ref{fig:bfd1100} represent the oscillatory solutions obtained from DNS.

Note that, we also have stable limit cycles at the onset of convection due to the supercritical Hopf bifurcation at $r = 1$. These limit cycles grow in size with the increment in $r$ which cause a little enhancement in heat transport. However,  we do not observe any hysteresis during the backward continuation in this case and a scenario of  supercritical transition prevails at the onset. The limit cycle vanishes for $r \geq 1.22$ and we get the high amplitude 2D rolls solution for subsequent higher values of $r$. 
To understand the underlying reason, we calculate  the basins of attraction for different solutions. Figure~\ref{fig:basin_attraction} shows  the basins of attraction for the 2D rolls solution (white region) and the periodic oscillatory rolls solution (black region) corresponding to four different values of $r$. From the figure, we see that the basin of attraction for limit cycles shrinks as $r$ is increased. Simultaneously, the limit cycle increases in size and becomes homoclinic to the co-existing 2D rolls saddle at $r \sim 1.22$ and ceased to exist thereafter. The projection of these limit cycles, their distance from the 2D rolls saddle and the time period of oscillation are shown in the figure~\ref{fig:homoclinic_bif}. The homoclinic bifurcation of the limit cycle is apparent from the figure. Therefore, for $r > 1.22$, the finite amplitude solution which originated from the 2D rolls branch is observed in the model. Similar qualitative behaviour in the dynamics is also observed in DNS. 
\begin{figure}[h]
\includegraphics[height=3in, width = 0.5\textwidth]{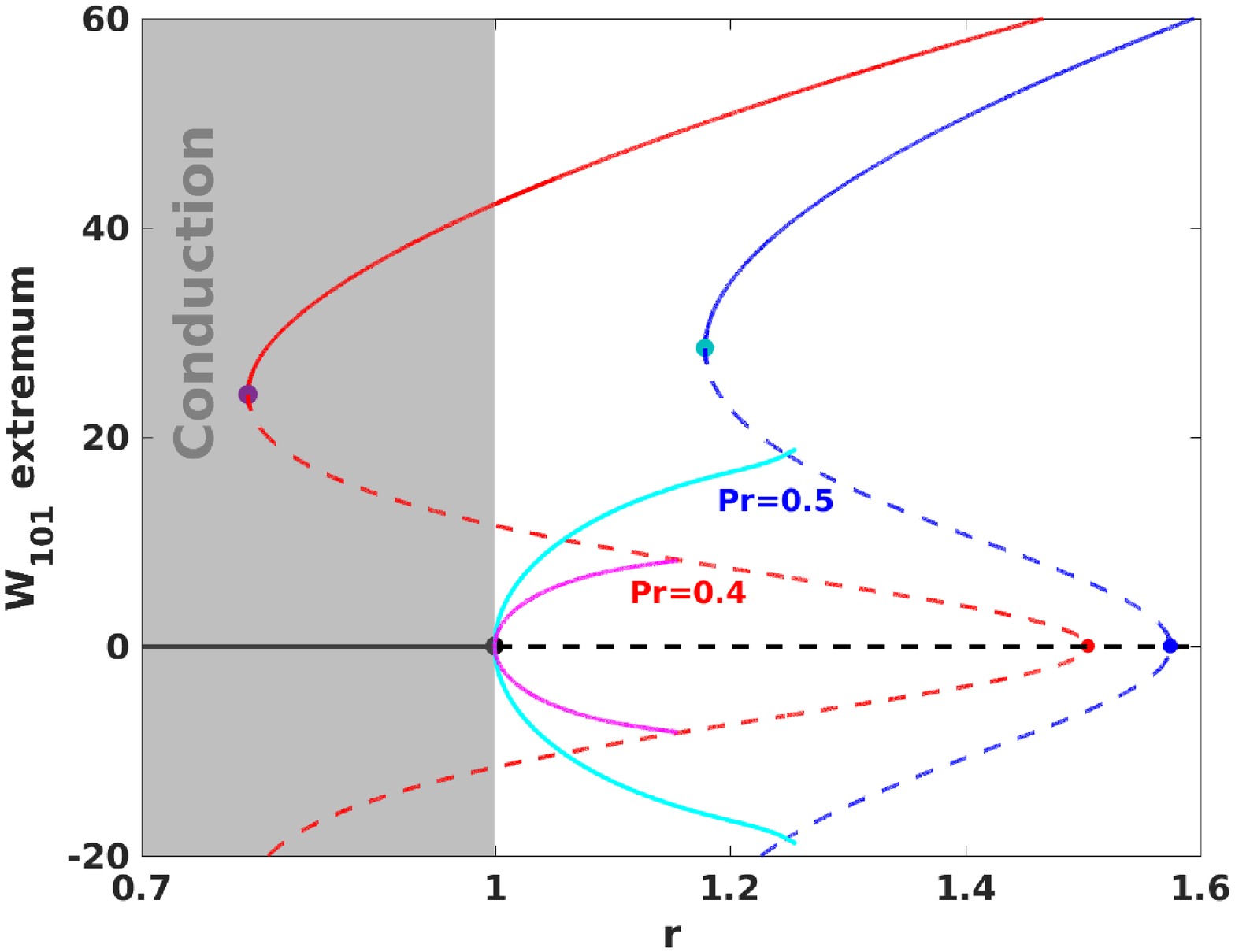}
\caption{Bifurcation diagram is constructed using the model for $\mathrm{Q}=~100$ at two different values of $\mathrm{Pr}$. Solid and dashed curves, respectively, represent the stable and unstable solutions. The blue and cyan curves, respectively, represent steady 2D rolls and oscillatory rolls solutions for $\mathrm{Pr}=~0.5$, while the red and pink curves represent the same for $\mathrm{Pr}~=~0.4$. It is evident that the subcritical bifurcation point moves away from the onset of convection as $\mathrm{Pr}$ is increased from $0.4~\mathrm{to}~0.5$.}
\label{fig:pr_effect}
\end{figure}
\begin{figure}[h]
\includegraphics[height=!, width = 0.5\textwidth]{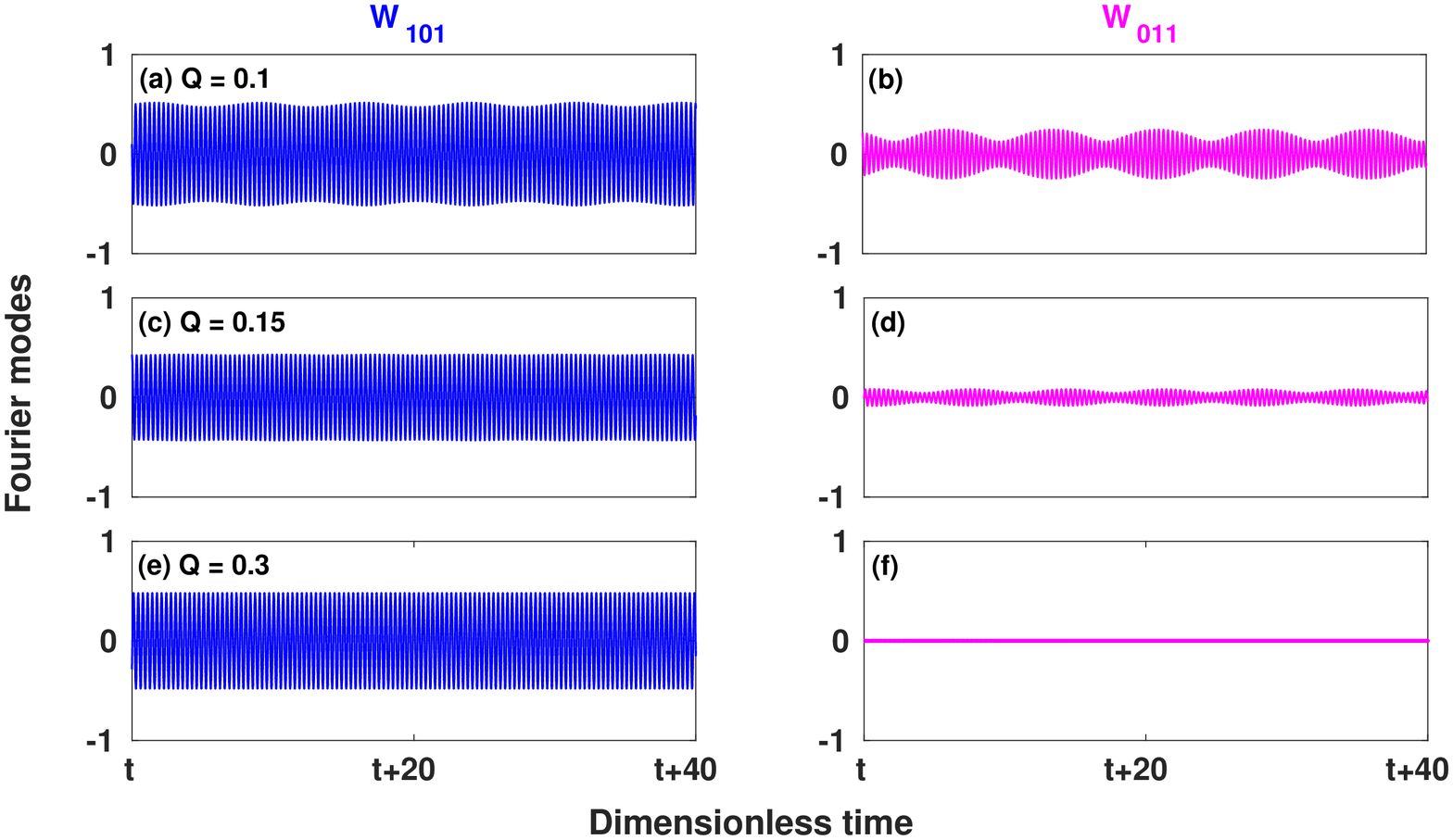}
\caption{Temporal evolution of the Fourier modes $W_{101}$ (blue) and $W_{011}$ (pink) near the convection onset ($r=1.004$) for $\mathrm{Ta}=1100$, $\mathrm{Pr} = 0.1$ and three different values of $\mathrm{Q}$.}
\label{fig:effect_of_Q}
\end{figure}

As we mentioned earlier, changes in $\mathrm{Ta}$ corresponding to a fixed $\mathrm{Pr}$ do not affect the bifurcation scenario near the onset of convection qualitatively. However, the width of the hysteresis loop decreases by a small amount with the increment in $\mathrm{Ta}$. This causes a diminution in $\mathrm{Nu}$ near the onset of convection which we have seen earlier in Section~\ref{DNS_result}~(see figure~\ref{fig:onset_nu}). The bifurcation scenario also remains unchanged with the variation in $\mathrm{Q}$ for fixed $\mathrm{Ta}$ and $\mathrm{Pr}$. This can be easily verified by observing that the low dimensional model does not contain any term related to $\mathrm{Q}$. 

The bifurcation scenario near the onset of convection becomes qualitatively different as $\mathrm{Pr}$ is varied in our considered range. As $\mathrm{Pr}$ increases, the saddle node bifurcation point moves towards the Hopf bifurcation point. As a result, the width of the hysteresis loop gradually decreases with the increment in $\mathrm{Pr}$. The saddle node bifurcation point eventually leaves the conduction region for $\mathrm{Pr} = 0.46$ and the scenario of  subcritical transition vanishes there. However, the 2D rolls branch with subcritical origin exists,  but it turns around ahead of the conduction region. So,  the scenario of  supercritical transition prevails at the onset due to the supercritical Hopf bifurcation. Figure~\ref{fig:pr_effect} displays the scenario corresponding to two different values of $\mathrm{Pr}$.  It is clearly seen from the figure that the scenario of  subcritical transition persists at the onset for $\mathrm{Pr} = 0.4$ while it vanishes for $\mathrm{Pr} = 0.5$. Further increment in $\mathrm{Pr}$ eliminates the possibility of overstability as discussed earlier. 

\subsection{Effect of small magnetic field ($\mathrm{Q} < 100$)}\label{small_Q}
 We now discuss the results of DNS performed near the onset of overstable rotating convection in the presence of a weak magnetic field. Here we consider two different values of the Taylor number ($\mathrm{Ta} = 1100, 3000$) for $\mathrm{Pr} = 0.1$  and vary $\mathrm{Q}$ in the range $0$ to $100$ and investigate the flow patterns close to the onset of convection.  

It has been reported earlier that in the absence of external magnetic field, when rotation acts solely, three dimensional (3D) quasiperiodic oscillatory cross-rolls ($W_{101}\neq0$, $W_{011}\neq0$, $max|W_{101}| = max|W_{011}|$) are observed at the onset of convection for smaller $\mathrm{Ta}$, while 3D periodic oscillatory rolls ($W_{101}\neq0$, $W_{011} = 0$) are observed for higher $\mathrm{Ta}$~\cite{hirdesh:2013}. The presence of magnetic field in the horizontal direction breaks the $x \rightleftharpoons y$ symmetry of the system. As a result, the quasiperiodic cross-rolls with equal amplitudes which were observed corresponding to the lower values of $\mathrm{Ta}$ in the absence of magnetic field now become asymmetric ($max|W_{101}| \neq max|W_{011}|$) in nature and the amplitude of the Fourier mode $W_{011}$ becomes smaller compared to that of $W_{101}$ (see Figure~\ref{fig:effect_of_Q}). From the Figure~\ref{fig:effect_of_Q}(a), it is prominent that even the presence of a very weak magnetic field ($\mathrm{Q} = 0.1$) makes the flow asymmetric. A little increment in $\mathrm{Q}$ causes further diminution in the amplitude of $W_{011}$ (see Figure~\ref{fig:effect_of_Q}(b)) and eventually suppresses its oscillation for $\mathrm{Q} = 0.3$ (see Figure~\ref{fig:effect_of_Q}(c)). As a result, 2D oscillatory rolls for which $W_{011} = 0$ are observed at the onset.

\begin{figure}[h]
\includegraphics[height=!, width = 0.5\textwidth]{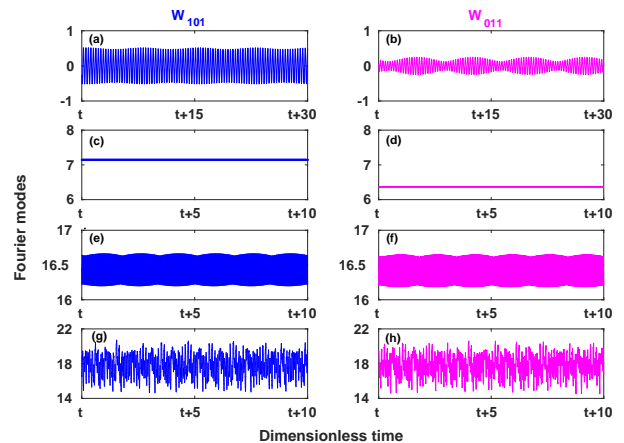}
\caption{Temporal variation of the Fourier modes $\mathrm{W_{101}}$ (magenta) and $\mathrm{W_{011}}$ (pink) for $\mathrm{Ta}= 1100$, $\mathrm{Q} = 0.1$, and $\mathrm{Pr}=0.1$ for four different values of $r$. The reduced Rayleigh number $r$ for the first row (a)-(b), second row (c)-(d), third row (e)-(f) and the last row (g)-(h) are, respectively, $1.004$, $1.029$, $1.479$ and $2.017$.}
\label{fig:small_Q_ts}
\end{figure}

We also vary $r$ for fixed $\mathrm{Ta} = 1100$ and $\mathrm{Q} = 0.1$ to investigate the subsequent transitions in overstable rotating convection in the presence of a very weak horizontal magnetic field after overstability sets in. Figure~\ref{fig:small_Q_ts} shows the variation of the two largest Fourier modes $W_{101}$ and $W_{011}$ corresponding to the transitions that occur following overstable onset in an RMC system as the value of $r$ is increased.  We observe asymmetric quasiperiodic cross rolls at the overstability onset~(see fig.~\ref{fig:small_Q_ts}(a)) followed by cross rolls~(see fig.~\ref{fig:small_Q_ts}(b)) for which $|W_{101}| \neq |W_{011}|$ and $|W_{101}| > |W_{011}|$ as $r$ is increased. Further increment in $r$ exhibits quasiperiodic cross rolls~(see fig.~\ref{fig:small_Q_ts}(c)) for $r = 1.479$. Finally, chaotic cross rolls~(see fig.~\ref{fig:small_Q_ts}(d)) appear as we raise the value of $r$ further. 

\begin{figure}[h]
\includegraphics[height=!, width = 0.5\textwidth]{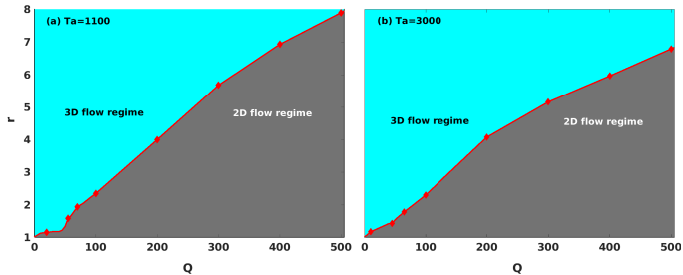}
\caption{Flow regimes on two parameter $\mathrm{Q} - r$ plane for $\mathrm{Pr} = 0.1$ and two different values of $\mathrm{Ta}$ as obtained from DNS. The gray regions in both figures show the 2D flow regimes, while the cyan regions represent the 3D flow regimes. From both figures, it is apparent that the 2D flow regime shrinks with the increment in $\mathrm{Ta}$.}
\label{fig:two_par_2D_3D_transition}
\end{figure}

Now, we construct two diagrams from DNS data to show the 2D and 3D flow regimes on the $\mathrm{Q} - r$ plane corresponding to two different values of $\mathrm{Ta}$ (see Figure~\ref{fig:two_par_2D_3D_transition}). The 2D flow regimes include both oscillatory and stationary rolls. For lower $\mathrm{Q}$ ($< 40$ and $<20$ for $\mathrm{Ta} = 1100$ and $3000$ respectively), 2D flow regimes only include the oscillatory solutions and as soon as $r$ is raised a little beyond the onset of convection, the flow becomes three dimensional. The bifurcation structure associated with these three dimensional flow patterns are found to be similar to the ones reported in Ref.~\cite{hirdesh:2013} in rotating convection. However, for higher $\mathrm{Q}$, in the 2D flow regime both oscillatory and stationary flow patterns coexist. The bifurcation structures associated with these co-existing flow patterns have already been discussed in the Section~\ref{large_Q}.

\section{Conclusions}\label{sec:conclusions}
We have investigated the effect of external uniform horizontal magnetic field on overstable rotating convection using Rayleigh-B\'{e}nard geometry of electrically conducting low Prandtl number fluids with stress free boundary conditions. A combination of linear stability analysis, three dimensional (3D) direct numerical simulations (DNS) and low dimensional modeling of the system is performed for this purpose.  The parameters $\mathrm{Ta}$, $\mathrm{Q}$, and $\mathrm{Pr}$ are respectively varied in the ranges $750 \leq \mathrm{Ta} \leq 3000$, $0 < \mathrm{Q} \leq 1000$, and $0 < \mathrm{Pr} \leq 0.5$. 

Linear analysis of the system reveals that in this parameter regime, 2D rolls aligned along the magnetic field are the preferred mode of convection which is found to saturate to two dimensional oscillatory rolls. Interestingly, a finite amplitude steady rolls solution is found to coexist with the oscillatory rolls solution at the onset of convection for relatively larger values of $\mathrm{Q}$. The presence of the finite amplitude steady solution also causes much higher heat transport near the onset compared to that of the oscillatory solution. Surprisingly, the flow remains two dimensional even in the nonlinear regime for $\mathrm{Q} \sim 100$. The regime of two dimensionality enhances with the increment in $\mathrm{Q}$. Even, a weak magnetic field ($\sim 0.5$) is found to be sufficient to  maintain  two dimensionality in the nonlinear regime. This behavior is different from the case when the imposed magnetic field is vertical, which has been much investigated experimentally as well as theoretically. This suggests that an inclined field with even a small horizontal component might render the flow two-dimensional, but this needs to be confirmed with further work.

A convenient three mode model is derived from the DNS data to uncover the bifurcation structure associated with the two dimensional flow patterns for larger $\mathrm{Q} ~(\geq 100)$. Analysis of the model along with the performance of DNS clearly establishes the simultaneous presence of subcritical and supercritical branches of convection in a wide range of the parameter space.  Bifurcation analysis of the model also reveals that the appearance of finite amplitude solutions at the onset is associated with a subcritical steady rolls branch generated through a subcritical pitchfork bifurcation of the unstable conduction solution. This subcritical branch exists at Rayleigh numbers well below the critical for onset. Changes in Prandtl number ($\mathrm{Pr}$) also affect the scenario of transition to convection deeply. The scenario of subcritical transition disappears from the system as the value of $\mathrm{Pr}$ is increased. As a result, only the suprecritical transition to convection exists there.

\section{Acknowledgments} P.P. acknowledges support from Science and Engineering Research Board (Department of Science and Technology, India) (Grant No. MTR/2017/000945). M.G. is supported by INSPIRE programme of DST, India (Code: IF150261). Authors thank Paromita Ghosh, Lekha Sharma and Sutapa Mandal for their fruitful comments.

\vskip2pc

%\bibliographystyle{apsrev4-1} %%%% .BST file

%\bibliography{manojit_rmc} %%%%% .Bib file
%merlin.mbs apsrev4-1.bst 2010-07-25 4.21a (PWD, AO, DPC) hacked
%Control: key (0)
%Control: author (72) initials jnrlst
%Control: editor formatted (1) identically to author
%Control: production of article title (-1) disabled
%Control: page (0) single
%Control: year (1) truncated
%Control: production of eprint (0) enabled
%

\end{document}